\documentstyle[11pt,newpasp,twoside,epsf]{article}
\def\edcomment#1{\iffalse\marginpar{\raggedright\sl#1\/}\else\relax\fi}
\reversemarginpar

\begin{document}
\title{Studying the Gaseous Phases of Galaxies with Background QSOs}
\author{Jane C. Charlton}
\affil{Department of Astronomy \& Astrophysics, The Pennsylvania
State University, 525 Davey Laboratory, University Park, PA 16802}

\begin{abstract}
High resolution rest frame UV quasar absorption spectra covering low
and high ionization species, as well as the Lyman series lines,
provide remarkably detailed information about the gaseous phases of
galaxies and their environments.  For redshifts less than 1.5, many
important chemical transitions remain in the observed ultraviolet
wavelength range.  I present examples of absorption that arises from
lines of sight through a variety of structures, drawn from UV spectra
recently obtained with STIS/HST.  Even with the greater sensitivity of
COS/HST there will be a limit to how many systems can be studied in detail.
However, there is a great variety in the morphology of the phases of gas
that we observe, even passing through different regions of the same
galaxy.  In order to compile a fair sample of the gaseous structures
present during every epoch of cosmic history, hundreds of systems must
be sampled.  Multiple lines of sight through the same structures are
needed, as well as some probing nearby structures whose luminous hosts
have been studied with more standard techniques.  Combined with high
resolution optical and near--IR ground--based spectra, it will be
possible to uniformly study the gaseous morphologies of galaxies of
all types through their entire evolutionary histories.
\end{abstract}

\section{Introduction}

The tool of quasar absorption line spectroscopy has several distinct
advantages over more traditional imaging studies of distant galaxies.
Spectra covering absorption lines from numerous chemical elements in
various states of ionization can yield detailed information about the
physical conditions in the gaseous components of the universe.  This
is not limited to only the most luminous components.  Dwarf galaxies,
low surface brightness galaxies, and even intergalactic structures are
probed by quasar lines of sight.  Quasar absorption lines can be used
to study gas during the birth and death of stars and of entire
galaxies.  The kinematic information contained in high resolution
spectra allows us to study processes.  This is not just a still
picture snapshot; it is more like a short movie.  Finally, the same
level of detail is available for our study at all redshifts because
the same method can be applied using optical and near--IR spectroscopy.

However, the study of quasar absorption lines also presents some
challenges if we are to extract from the quasar spectra all of
the detailed information about physical conditions of gas.
Imagine if we had to classify a galaxy according to its standard
morphological type by zooming in on an image of just a small part
of the galaxy.  We have to consider carefully what a single line
of sight tells us about the global conditions in galaxies.
In fact, it should be possible to learn a great deal if we consider
the evolution of the ensemble of gaseous structures probed by quasar
lines of sight.  In order to realize this potential, we must learn
to connect absorption signatures to the physical conditions of the
phases of gas that produce them, and to the processes that give
rise to such signatures in local galaxies.

For the optimal study of the detailed physical conditions along a
quasar line of sight we need high resolution spectroscopy covering all
rest frame UV transitions.  For redshifts less than one, a significant
fraction of the key transitions still appear in the ultraviolet.
In this proceedings, I review examples of various systems that have
been studied in detail, focusing on the question of what additional
data would yield a significant advance.  The goal is to define
capabilities for the optimal UV spectrograph and telescope to be used
for such a program.

\begin{figure}
\plotone{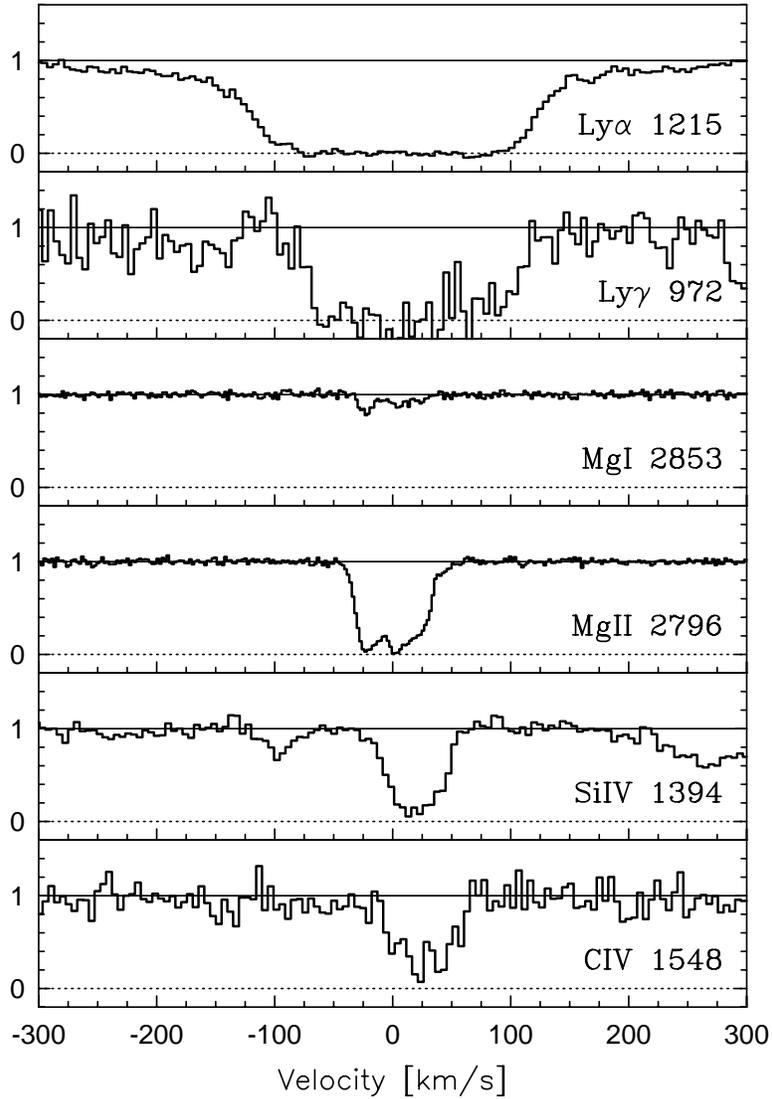} \caption {Selected transitions for
the $z=0.9902$ strong \hbox{{\rm Mg}\kern 0.1em{\sc ii}} system along
the PG~$1634+706$ line of sight, presented in velocity space.  The
velocity zero--point corresponds to the apparent optical depth
centroid for the \hbox{{\rm Mg}\kern 0.1em{\sc ii}} profile.  The
\hbox{{\rm Mg}\kern 0.1em{\sc i}} and \hbox{{\rm Mg}\kern 0.1em{\sc
ii}} profiles were obtained with HIRES/Keck, with $R=45,000$
(P.I. Churchill), and the other transitions were observed with
STIS/HST, with $R=30,000$ (P.I.'s Jannuzi and Burles).}
\end{figure}

\section{A \hbox{{\rm C}\kern 0.1em{\sc iv}}--Deficient Strong \hbox{{\rm Mg}\kern 0.1em{\sc ii}} Absorber?}

The $z=0.99$ absorber toward the quasar PG1634+706 is a strong
\hbox{{\rm Mg}\kern 0.1em{\sc ii}} absorber, which implies that it is
very likely to be within an impact parameter of $35h^{-1}$~kpc of an
$\sim L^*$ galaxy (Steidel 1995).  It is part of the small subset of
\hbox{{\rm Mg}\kern 0.1em{\sc ii}} absorbers that have particularly
small \hbox{{\rm C}\kern 0.1em{\sc iv}} absorption features, i.e.  it
is \hbox{{\rm C}\kern 0.1em{\sc iv}}--deficient relative to other
strong \hbox{{\rm Mg}\kern 0.1em{\sc ii}} absorbers
(Churchill et~al. 2000).  STIS/E230M spectra of this quasar have been
obtained by Jannuzi and by Burles, and HIRES/Keck spectra (covering
\hbox{{\rm Mg}\kern 0.1em{\sc i}}, \hbox{{\rm Mg}\kern 0.1em{\sc ii}}, 
and \hbox{{\rm Fe}\kern 0.1em{\sc ii}} ) have been obtained by
Churchill and Vogt (2001).  The most important transitions are shown in
Figure 1.  Detailed results of our modeling of this system have been
presented by Ding et~al. (2002).

The system can be described by a combination of a minimum of four
different phases of gas.  A phase is a region or regions with density
and temperature within some range, spatially separate from other
phases.  The \hbox{{\rm Mg}\kern 0.1em{\sc ii}} absorption arises in
clouds that have densities of $\sim 0.01$~\hbox{cm$^{-3}$}.  The bulk
of the \hbox{{\rm Mg}\kern 0.1em{\sc i}} absorption cannot arise in
the same clouds.  Instead, we propose that the \hbox{{\rm Mg}\kern
0.1em{\sc i}} is produced by small ($\sim 100$~AU), cool pockets of
the interstellar medium.  The metallicity of the \hbox{{\rm Mg}\kern
0.1em{\sc ii}} and \hbox{{\rm Mg}\kern 0.1em{\sc i}} clouds is about
$0.2\times$solar.  A ``broad'', low-metallicity ($<0.01\times$solar)
component is required to self--consistently fit \hbox{{\rm Ly}\kern
0.1em$\alpha$}, \hbox{{\rm Ly}\kern 0.1em$\beta$}, and the
higher--order Lyman--series lines.  The low metallicity could relate
to why this absorber is \hbox{{\rm C}\kern 0.1em{\sc iv}} deficient.
Finally, a strong, smooth \hbox{{\rm Si}\kern 0.1em{\sc iv}}
absorption profile suggests an additional collisionally ionized phase
with $T\sim 60,000$~K, perhaps heated by shocks.

For this system, there are two outstanding issues of particular
interest.  First, with higher resolution ($R>100,000$)
we could test the narrow \hbox{{\rm Mg}\kern 0.1em{\sc i}} cloud
hypothesis.  Second, an image of the host galaxy (or those for similar
systems) would enable us to see whether an early--type galaxy is
responsible for the lack of absorption resembling the Milky Way
corona.

\section{A Redshift One Galaxy Group?}

The $z\sim0.93$ systems in the PG~$1206+459$ line of sight are the
subject of a separate contribution in this volume by Jie Ding, and
therefore no detail or figures will be presented here.  Three distinct
systems (two strong \hbox{{\rm Mg}\kern 0.1em{\sc ii}} and one
weak \hbox{{\rm Mg}\kern 0.1em{\sc ii}}) are apparent at
$z=0.925$, $z=0.928$, and $z=0.934$, an overall velocity spread of
just over $1000$~\hbox{km~s$^{-1}$}.  This is likely to be a group of
galaxies, with the three systems arising in three very different types
of galaxy hosts.  The study of these systems would be aided by higher
$S/N$ since there are suggestions of abundance pattern variations.
Understanding of the phases giving rise to high ionization absorption
would be much better constrained if a high resolution spectrum
covering the
\hbox{{\rm O}\kern 0.1em{\sc vi}} transition was obtained.

\begin{figure}
\plotone{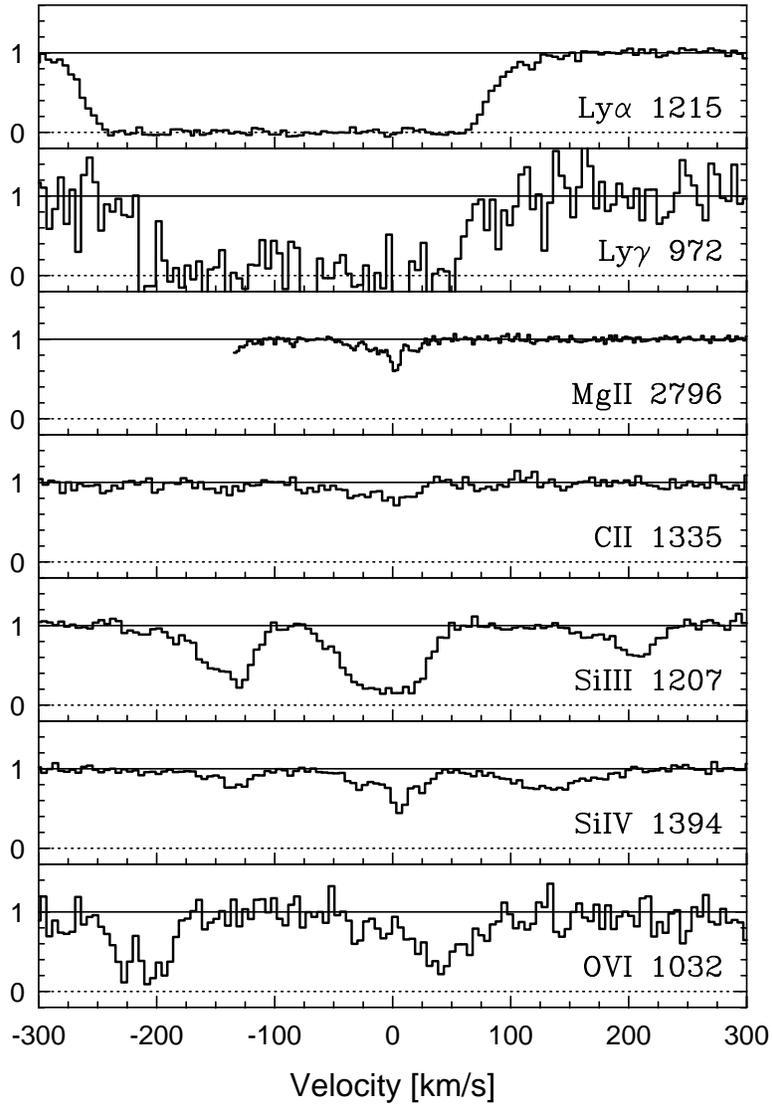} \caption {Selected transitions for the
$z=1.0414$ multiple--cloud, weak \hbox{{\rm Mg}\kern 0.1em{\sc ii}}
system along the PG~$1634+706$ line of sight, presented in velocity
space.  These are taken from the same spectra noted in the Figure 1
caption.  }
\end{figure}

\section{A Pair of Dwarf Galaxies?}

The $z=1.04$ Absorber Toward PG~$1634+706$ is a multiple cloud, weak
\hbox{{\rm Mg}\kern 0.1em{\sc ii}} absorber, with two sets (kinematic
subsystems) of low ionization ``clouds'' separated by $\sim
150$~\hbox{km~s$^{-1}$}.  Each of the subsystems has an \hbox{{\rm
O}\kern 0.1em{\sc vi}} absorption profile offset by $\sim
50$~\hbox{km~s$^{-1}$}.  Profiles for key transitions from the
STIS/HST and HIRES/Keck spectra are shown in Figure 2.  This system
was modeled by Zonak et~al. (2002), and the results are summarized
here.

The \hbox{{\rm Mg}\kern 0.1em{\sc ii}} clouds have metallicity
$<0.03\times$solar, constrained by a partial Lyman limit break.  Although
\hbox{{\rm Si}\kern 0.1em{\sc iv}} has the same kinematic structure as
\hbox{{\rm Mg}\kern 0.1em{\sc ii}} in the stronger subsystem, the
clouds appear to be offset by velocities ranging from
$3$--$5$~\hbox{km~s$^{-1}$}.  This could be related to an \hbox{{\rm
H}\kern 0.1em{\sc ii}} region flow, but the interpretation of the
multiple cloud structure remains unclear.  The \hbox{{\rm Si}\kern
0.1em{\sc iii}} profile is smooth, but unsaturated, and is stronger
than can be explained by the combination of the \hbox{{\rm Mg}\kern
0.1em{\sc ii}} and \hbox{{\rm Si}\kern 0.1em{\sc iv}} cloud phases.
This suggests a collisional ionization mechanism for producing
\hbox{{\rm Si}\kern 0.1em{\sc iii}}.  The \hbox{{\rm Ly}\kern
0.1em$\alpha$} profile requires two additional photoionized phases
that can also produce the broad \hbox{{\rm O}\kern 0.1em{\sc vi}}
profiles.

The low metallicity and the kinematics of the two subsystems
suggest that two dwarf galaxies and their halos could be responsible.
The lower ionization phases would arise in the inner regions of
the dwarfs, and the offset higher ionization phases from their halos.

Very little is known about the expected absorption signature
of dwarf galaxies.  If quasar absorption line probes through
nearby dwarfs were available, this mystery would be quickly
solved.  However, this would require that some fainter quasars
were accessible for high resolution spectroscopy.

\begin{figure}
\plotone{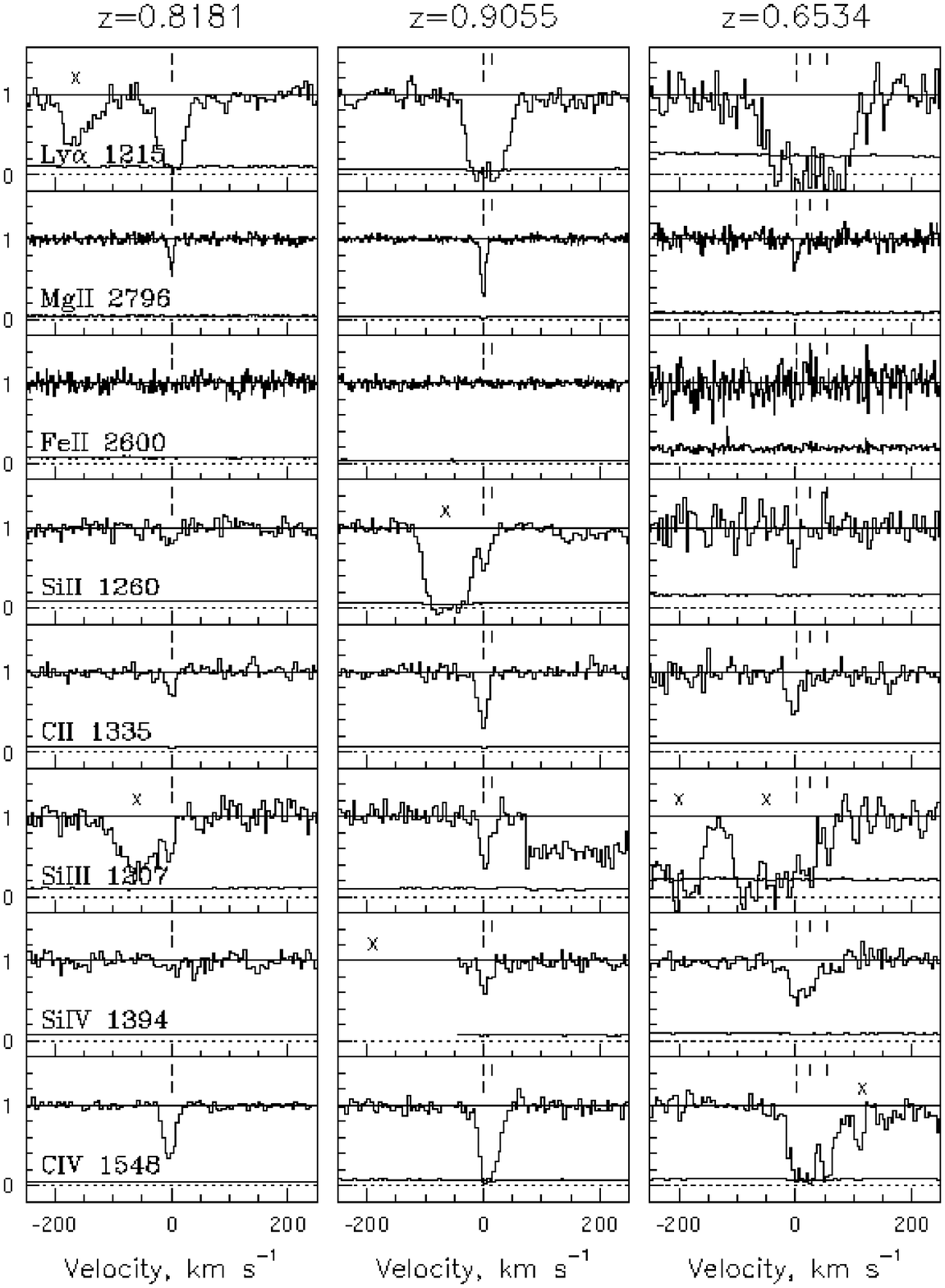} \caption {A comparison of selected transitions
for the three single--cloud, weak \hbox{{\rm Mg}\kern 0.1em{\sc ii}}
absorbers along the PG$1634+706$ line of sight.  The spectra from
which these were extracted are referenced in the Figure 1 caption.  }
\end{figure}

\section{A Variety of Weak \hbox{{\rm Mg}\kern 0.1em{\sc ii}} Absorbers}

Figure 3 shows selected transitions for the three single--cloud, weak
\hbox{{\rm Mg}\kern 0.1em{\sc ii}} absorbers at $z=0.65$, $z=0.81$,
and $z=0.90$ along the same quasar PG~$1634+706$ line of sight.
Unlike the strong \hbox{{\rm Mg}\kern 0.1em{\sc ii}} absorbers,
single--cloud, weak \hbox{{\rm Mg}\kern 0.1em{\sc ii}} absorbers are
generally not found within $50$~kpc of $L^*$ galaxies.  Rigby,
Charlton, \& Churchill (2002) found, based on a larger sample for
which lower resolution FOS spectra were available, that they cannot be
far below solar metallicity.  They also found that at least a subset
of these absorbers are very small in dimension.  Detailed models of
the three PG~$1634+706$ single--cloud, weak \hbox{{\rm Mg}\kern
0.1em{\sc ii}} absorbers were presented in Charlton et~al. (2002).

The $z=0.81$ weak \hbox{{\rm Mg}\kern 0.1em{\sc ii}} absorber is the
kinematically simplest along this line of sight.  The \hbox{{\rm
C}\kern 0.1em{\sc iv}} absorption is relatively weak.  It was not even
detected in a low resolution FOS spectrum.  The \hbox{{\rm Ly}\kern
0.1em$\alpha$} profile constrains the metallicity to be supersolar,
and the ionization conditions indicate that the low ionization phase
is sub--parsec scale.  Despite the \hbox{{\rm C}\kern 0.1em{\sc iv}}
being weak, the broad shape of its profile cannot be fit by the
\hbox{{\rm Mg}\kern 0.1em{\sc ii}} cloud.  The high ionization phase
could be collisionally ionized, but photoionization is more likely,
with a cloud size of $\sim 100$~pc.  The high ionization phase could
surround or be surrounded by the low ionization gas.

If the high ionization material is a thicker shell surrounding a small
low ionization pocket then we would expect to see many systems with
only the high ionization phase.  A large statistical sample is needed
to determine the geometry.  To distinguish between collisional
ionization and photoionization, coverage of the \hbox{{\rm O}\kern
0.1em{\sc vi}} transition is key.

In contrast, the $z=0.90$ weak \hbox{{\rm Mg}\kern 0.1em{\sc ii}}
absorber has relatively strong \hbox{{\rm C}\kern 0.1em{\sc iv}}.  A
low ionization phase has supersolar metallicity and a cloud size of
$\sim 50$~pc.  The \hbox{{\rm C}\kern 0.1em{\sc iv}} and \hbox{{\rm
N}\kern 0.1em{\sc v}} require a separate, broader cloud, which is
consistent with photoionization and not with collisional ionization.
Also, in this case, an additional offset high ionization cloud is
required to fit the blue wing of the \hbox{{\rm C}\kern 0.1em{\sc iv}}
and the \hbox{{\rm Ly}\kern 0.1em$\alpha$}, which is not centered on
the \hbox{{\rm Mg}\kern 0.1em{\sc ii}}.  The \hbox{{\rm C}\kern
0.1em{\sc iv}} equivalent width could be large in this case partly
because the line of sight passed through a separate cloud at a
different velocity, and not because the centered \hbox{{\rm C}\kern
0.1em{\sc iv}} cloud has special physical conditions.

The \hbox{{\rm C}\kern 0.1em{\sc iv}} profile associated with the
$z=0.65$ weak \hbox{{\rm Mg}\kern 0.1em{\sc ii}} absorber has distinct
subcomponent structure, and therefore a large equivalent width.  The
\hbox{{\rm Ly}\kern 0.1em$\alpha$} is also quite strong for this
system, possibly also because of the large kinematic spread.  It is
centered on the \hbox{{\rm C}\kern 0.1em{\sc iv}} and not on the
single \hbox{{\rm Mg}\kern 0.1em{\sc ii}} component.  The low
ionization phase is poorly constrained, but its metallicity must
exceed $0.1\times$solar.  The three \hbox{{\rm C}\kern 0.1em{\sc iv}}
components arise in high ionization, kiloparsec--scale structures,
which are likely to be photoionized.

What are these weak, single--cloud \hbox{{\rm Mg}\kern 0.1em{\sc ii}}
absorbers?  The small size constraints on the low ionization phases
implies that there must be huge numbers of the host structures.  In
fact, to account for the observed number of weak, single--cloud
\hbox{{\rm Mg}\kern 0.1em{\sc ii}} absorbers, per unit redshift, there
must be more than a million hosts for every $L^*$ galaxy in the
universe.  Yet, these are not associated with $L^*$ galaxies.  They
appear to be self--enriched pockets of higher density material
embedded in higher ionization, more diffuse gas.  Rigby et~al. (2002)
discuss the origin of the weak
\hbox{{\rm Mg}\kern 0.1em{\sc ii}} absorbers at
length.  Their two most likely explanations are small traces of gas
in fading remnants of Population III star clusters and fragments of
supernova shells in faint blue dwarf galaxies.

A number of new capabilities would help to constrain the properties of
this intriguing class of object.  Double line of sight observations at
a scale of parsecs, though they could not be done for many systems,
would provide an essential check on size constraints for these weak
\hbox{{\rm Mg}\kern 0.1em{\sc ii}} absorbers.  Nearby weak
\hbox{{\rm Mg}\kern 0.1em{\sc ii}} absorbers must be identified
so that narrow band imaging can be used to discover if hosts are dwarf
galaxies or isolated star clusters.  Of course, a large statistical
sample would clarify whether there are different classes of weak,
single--cloud \hbox{{\rm Mg}\kern 0.1em{\sc ii}} absorbers, with
different types of hosts.

\section{Conclusions}

The small sample of systems presented here demonstrates great promise
for learning about the detailed physical conditions in a variety of
gaseous environments.  For these systems, the data have some
limitations and, as a result, some questions remain unanswered.
These highlight the need for additional UV spectroscopy capabilities.
I conclude by presenting a ``wish list'' if we aim to construct the
dynamical history of the ensemble of galaxies and gaseous structures:

\begin{itemize}
\item{hundreds of systems, or even thousands!}
\item{cover all key chemical species from the rest frame ultraviolet}
\item{study many systems that are at low enough redshift to image
their galaxy hosts; these serve as a calibrator for higher redshift
systems}
\item{higher spectral resolution ($R=100,000$ or even higher) for a
subset of the systems in order to resolve interstellar medium features}
\item{high signal--to--noise ($>20$) for a subset to enable abundance
pattern studies}
\item{multiple lines of sight through the same objects}
\item{a separate, detailed study of the interstellar medium of the
Milky Way and of nearby galaxies using the same techniques}
\end{itemize}

Support for this work was provided by the NSF (AST--9617185) and by NASA
(NAG 5--6399 and HST--GO--08672.01--A), the latter from the
Space Telescope Science Institute,
which is operated by AURA, Inc., under NASA contract NAS5--26555.


\begin{references}
\reference{Charlton, J. C., Ding, J., Zonak, S. G., Churchill, C. W.,
\& Bond, N. A. 2002, \apj, submitted}
\reference{Churchill, C. W.,  Mellon, R. R., Charlton, J. C., Jannuzi, B. T.,
Kirhakos, S., \& Steidel, C. C., 2000, \apj, 543, 577}
\reference{Churchill, C. W., \& Vogt, S. S. 2001, \aj, 122, 679}
\reference{Ding, J., Charlton, J. C., Bond, N. A., Zonak, S. G.,
\& Churchill, C. W. 2002, \apj, submitted}
\reference{Ding, J., Charlton, J. C., Churchill, C. W., \& Palma, C. 2002,
in preparation}
\reference{Rigby, J. R., Charlton, J. C., \& Churchill, C. W. 2002,
\apj, 565, 743}
\reference{Steidel, C. C. 1995, in QSO Absorption Lines, ed. G. Meylan
(Garching:Springer--Verlag), 139}
\reference{Zonak, S. G., Charlton, J. C., Ding, J., \& Churchill, C. W. 2002,
\apj, submitted}
\end{references}
\end{document}